\documentclass[aps,prl,onecolumn,superscriptaddress,nofootinbib,nobalancelastpage,showpacs,nobibnotes]{revtex4}
\usepackage{epsfig}

\begin{document}
\newcommand{\identity}{\:\mbox{\sf 1} \hspace{-0.37em} \mbox{\sf 1}\,}

\title
{Absence of resonant enhancements in some inclusive rates}

\author{R. F. Sawyer}\email{sawyer@vulcan.physics.ucsb.edu}
\affiliation{Department of Physics, University of California at
Santa Barbara, Santa Barbara, California 93106}

\begin{abstract}

A toy model is defined and solved perturbatively with the aim
of examining some claimed ``resonant" enhancements of certain reaction
rates that enter popular models of leptogenesis. We find: a) that such enhancements
are absent; and b) that the perturbative solution, as done correctly using
finite-temperature field theory, is well defined without the ``resumming"
procedures found in the literature. The pathologies that led to the perceived
need for these procedures are an artifact of uncritical use of weighted
vacuum cross-sections in the determination of rates, without 
adequate attention to the effects of the medium upon the single particle
states within it.


\end{abstract}

\maketitle

\section{The model and its solution}

In some approaches to leptogenesis calculations based
on the assumed existence of heavy leptons, certain ``resonant" effects play a key role \cite{kw}-\cite{piffle2}.
In the present note we present a toy model in which, we believe, one can examine  
authoritatively some assumptions that are often made in the dynamical treatment
of these models, but which we shall find are false. We also intend our results
to serve as a cautionary warning against the uncritical calculation of reaction rates
in media by averaging vacuum cross-sections, calculated perturbatively, over 
unperturbed thermal distributions for the participating particles. 

We consider a set of particles described by:
\begin{enumerate}
\item  A scalar (Higgs) field, $\Phi$.
\item  A heavy neutral lepton field $N$.
\item  A light neutrino field $L$.
\item  A light ``anti-neutrino" field $\bar L$. 
\end{enumerate}

We can make our dynamical point most economically in a non-relativistic framework in which
the fields $N,L,\bar L$ all contain only annihilation (positive frequency) terms. If we were to
constructing sums of dangerous Feyman graphs, or those which enter the pathologies that we will be
discussing, there would be
a detailed correspondence between the graphs produced in the models of the literature and those produced
by the model below, where the particle annihilated by $\bar L$ is indeed the anti-particle of the particle 
annihilated by $L$.

The couplings that we shall include are those that enable the processes: $N\leftrightarrow L +\Phi$
with a strength parameter $g$ that is small enough to allow perturbative development in powers
of  $g$; and
$N\leftrightarrow \bar L +\Phi$ with a strength parameter $g'$ which is taken as very small. We seek results for the rate of  change of the number of $\bar L$'s, to lowest order $g'^2$. The 
interesting question is now how the interactions $N\leftrightarrow L +\Phi$, with strength $g$
perturb the lowest order result; the usual ``resonant" calculations find terms that are effectively
of lower order in $g$ than the superficial estimate $g'^2 g^2$. \footnote{To make clear the specific correspondence
of this remark with leptogenesis scenarios in the literature, we cite, e.g., two equations
from Ref. \cite{bari}; Eq.(13), in which superficial dependence of a relevant averaged cross-section,
is proportional to a factor $(h^\dagger h)^2$, where  $h$ is a coupling constant, and Eq.(18)
where the reaction rate (or the leading term thereof) is found to be of order $h^\dagger h$. Equivalent equations occur
in most of the references that we cite from this subject area.}

The physical context will be the following: without tracing the system back to its creation, I
take it to have evolved long enough, and to be subject to external influences that change slowly
enough, for the $N,L,\Phi$ system to have reached equilibrium. In contrast, the $\bar L$
sector is not in equilibrium; for simplicity we take the number of  $\bar L$'s to be zero at our
initial time.  The qualitative problem to be addressed 
originates in the fact that there are reactions
like $L+\Phi\rightarrow \bar L +\Phi$ in which the intermediate $N$ is unstable. This raises some familiar
questions about the treatment of unstable particles in field theory. But we would prefer to
call them ``familiar pseudo-questions", in view of our contention that thermal field theory
turns its crank in the same way, oblivious to whether particles are stable or not, as long
as we calculate rates that are sufficiently inclusive.

To exhibit the general structure of such processes requires only one particle in each reaction to have a distribution of energy
in its states. We take an $L$ or an $\bar L$ of momentum $\bf p_j$ to have energy $E_j$. The $L$'s will taken 
in a Fermi distribution,  $n_F(\epsilon_j)=[1+\exp (\beta \epsilon_j)]^{-1}$ where $\epsilon_j=E_j-\mu_L$.
We take the $\Phi$ particles all to have a common energy, $E_\Phi$. Taking the number of $\Phi$ states in a unit volume to be unity, the equilibrium
number density of $\Phi$'s is just the Bose function $n_B(\epsilon_\Phi)=[\exp (\beta \epsilon_\Phi)-1]^{-1}$
where $\epsilon_\Phi=E_\Phi-\mu_\Phi$. We take the $N$ particles to have the same number of states, but
with the Fermi distribution, $n_F(\epsilon_N)$, where  $\epsilon_N=E_N-\mu_N$. Finally we take the species
$N,\Phi,L$ to be in chemical equilibrium, $\mu_N=\mu_\Phi+\mu_L$. The $\bar L$'s will remain out of equilibrium.

For the $L$ and $\bar L$ fields we take the interaction picture representations,

\begin{eqnarray}
L(x,t)=(Vol)^{-1/2}\sum_{j}c_j e^{-i E_j t}e^{i {\bf p_j \cdot x}}
\nonumber\\
\bar L(x,t)=(Vol)^{-1/2}\sum_{j} \bar c_j e^{-i E_j \, t}e^{i {\bf p_j \cdot x}},
\end{eqnarray}
where $c_j$ and $\bar c_j$ are the respective annihilation operators for $L$ and $\bar L$.
Representing $N$ and $\Phi$, with annihilation operators, $a_i$ and $b_i$, respectively
we write the free Hamiltonian as,
\begin{equation}
H_0=E_N \sum_i a_i^\dagger a_i+E_\Phi  \sum_i b_i^\dagger b_i+\sum_j E_j(c_i^\dagger c_j+\bar c_j^\dagger \bar c_j).
\end{equation}

A local interaction of the fields $L$ and $\bar L$ with the fields $N$ and $\Phi$, where the latter two 
describe particles situated at the origin, is given by,  
\begin{equation}
H_I=g  \sum_j \Bigr [{f(E_j) \over \rho(E_j)} \Bigr ]^{1/2} (c_j+\epsilon ~ \bar c_j )[N ^\dagger \Phi ]_j +g\sum_j \Bigr [{f(E_j) \over \rho(E_j)} \Bigr ]^{1/2} (c_j^\dagger+\epsilon ~
\bar c_j^{\,\dagger})[N \Phi^\dagger ]_j\,\,\,\,,
\label{ham}
\end{equation}
where we have defined,

\begin{equation}
[N(t)^\dagger\, \Phi (t) ]_j=\int_V d^3x \,\,e^{-i{\bf  p_j \cdot x}} N^\dagger ({\bf x},t) \Phi ({\bf x},t) 
\end{equation}
\begin{equation}
[N(t)\, \Phi ^\dagger(t) ]_j=\int_V d^3x\,\, e^{i{\bf  p_j \cdot x}} N({\bf x,t}) \Phi^\dagger ({\bf x},t) 
\end{equation}
The function $f(E_j)$ in (\ref{ham}) includes the density-of-states factor $\rho(E_j)$, defined such that 
$\sum_k y(E_k)=\int dE_k\, \rho(E_k) y(E_k)$. It includes as well as any momentum dependence of the effective
coupling constant, and perhaps an ultra-violet cut-off factor. \footnote{ Of course all local, relativistic models will have vacuum processes that give ultraviolet divergences that need to be removed by renormalization. But these are irrelevant to
the issues we deal with here.}
 We choose parameters such that
$E_N-E_\Phi > E_j$ for some range of $L, \bar L$ states. The coupling that is ``lepton number violating"
in the sense that we defined above is given by $g' \equiv \epsilon g$. 

In our detailed example, in order to address a problem that is absolutely well defined,
we take the $N,\Phi, L$ to be in thermal and chemical equilibrium. We define $\mu_N,\mu_\Phi,
\mu_L$ as the chemical potentials for these species, with $\mu_N=\mu_\Phi+\mu_L$, and we define
the quantities, $\epsilon_N=E_N-\mu_N$ and $\epsilon_\Phi=E_\Phi-\mu_\Phi$. 
In the discussion section we shall return
to the relationship between the conditions assumed above and the somewhat different 
conditions that obtain in a typical leptogenesis calculation. But we emphasize that
our point is not to recapitulate these calculations, but rather to address in the simplest terms the 
dynamical questions raised below. 

Now we easily see the problem of  the Feynman graph approach. To first order
$g'^2$ and in the zero'th order in $g^2$ the rate of production of $\bar L$'s per unit volume,
where the initial state has no $\bar L$ 's is,
\begin{equation}
w^{(0)}=2 \pi g'^2n_F(\epsilon_N)[1+n_B(\epsilon_\Phi)]  f(E_N-E_\Phi),
\label{w0}
\end{equation}
where $n_F$ is the Fermi distribution, $n_F(\epsilon)=[1+\exp(\beta \epsilon)]^{-1}$,  $n_B$ is the Bose
distribution, $n_B(\epsilon)=[\exp(\beta \epsilon)-1]^{-1}$ and $\beta\equiv 1/T$.
To first order in $g^2$ we have two processes that produce $\bar L$'s: 

A. The process,  $\Phi+L \rightarrow N \rightarrow \Phi+\bar L$ with the rate
\begin{equation}
w^{(1)}= 2\pi g^2 g'^2 n_B(\epsilon_\Phi)[1+n_B(\epsilon_\Phi)]\int dE_i {n_F(E_i-\mu_L)\,\,f(E_i)^2  \over
(E_N-E_\Phi-E_i)^2}.
\label{gamma1}
\end{equation}

B. The process, $N+L \rightarrow \Phi +\bar L+L \rightarrow N+ \bar L$. For pedagogical simplicity we arrange 
the initial densities such that this process is insignificant compared to that of A. We point out in the subsequent
treatment exactly the way in which the perturbative anomalies in the rate for this second process get cancelled. 

The prevailing view in the literature is that the singularity in the integrand in (\ref{gamma1}) is regulated by a ``resummation" that makes the replacement,
\begin{equation}
\Bigr \vert { 1\over (E_N-E_\Phi -E_i) }\Bigr \vert ^2 \rightarrow \Bigr \vert { 1\over E_N-E_\Phi-E_i-i\gamma } \Bigr \vert ^2,
\label{replacement1}
\end{equation}
where $\gamma$ is the decay width of the $N$ particle, followed by another replacement,
\begin{equation}
\Bigr \vert {1\over E_N-E_\Phi -E_i-i\gamma } \Bigr \vert ^2=\Bigr \vert {E_N-E_\Phi-E_i +i \gamma \over (E_N-E_\Phi-E_i)^2+\gamma ^2} \Bigr \vert ^2
 \rightarrow \Bigr \vert { (E_N-E_\Phi-E_i)^2 \over (E_N-E_\Phi-E_i)^2+\gamma ^2 }\Bigr \vert ^2,
\label{replacement2}
\end{equation}
The first replacement, above, is the obvious one to
take into account the nature of an unstable particle in an intermediate state. The second replacement is supposed to remove a perceived double-counting, where a term in the the resonance production $\Gamma_1$ in the reaction 
$L+\Phi\rightarrow \bar L + \Phi$ has recapitulated the decay rate from the direct decay of an N.
We compute $\gamma$ from the decay rate for $N\rightarrow L+\Phi$
alone, and it is of order $g^2$, whence,

\begin{equation}
\int dE_i \, n_F(E_i) { f(E_i)^2 (E_N-E_\Phi-E_i)^2 \over
[(E_N-E_\Phi-E_i)^2+\gamma^2]^2 }=O(g^{-2})
\end{equation}
With this modification, $w^{(1)}$ of (\ref{gamma1}) is of order $g'^2 $, or reduced by two orders of $g$
from the superficial behavior of the graph, by virtue of the singularity regulation. We do not elaborate further on this procedure, which
we claim to be a plausible series of ansatzes, but incorrect. 

To find the correct expansion in powers of $g$ of the complete rate calculated to order $g'^2$, we show 
first that the field equations themselves give a simple formal expression, valid to order $g'^2$, for the net rate of change of
the number of $\bar L$'s (the analogue in our model of the rate of the change of lepton number 
in the theories which we are explicating). This expression is of the form of a Fourier transform of an expectation value
of a product two ``currents" evaluated at different space-time points, and is valid for \underline{any} 
configuration of the bystanding medium in which we are to take the expectation
value.

We define $Q=\sum_j  \bar c_j ^{\, \dagger} \bar{c_j}$ and note that
$Q$ would be conserved but for the $\epsilon g\equiv g'$ terms in H. We have
\begin{equation}
{dQ \over dt}=i [H',Q]=i(K_1-K_2),
\end{equation}
where
\begin{eqnarray}
K_1=g ' \sum_j \Bigr [{f(E_j) \over \rho(E_j)} \Bigr ]^{1/2} \bar c_j  [N ^\dagger \Phi ]_j
\nonumber\\
K_2=g'\sum_j \Bigr [{f(E_j) \over \rho(E_j)} \Bigr ]^{1/2}  \bar c_j^{\,\dagger} [N \phi ^\dagger]_j .
\end{eqnarray}
We wish to calculate $\langle dQ/dt \rangle$ where the bracket indicates the average value in
the medium. In the absence of the coupling $g'$ we choose the initial medium to be in an eigenstate
of $Q$ (we shall ultimately take $Q$ to be zero, as in the examples of (\ref{w0}) and (\ref{gamma1}), but this is not necessary.)
Then the expectation values of $K_1$ and $K_2$ are zero. The terms of order $g'^2$ in the expectation value for
$dQ/dt$ are generated by considering the linear response to the part of the Hamiltonian that is proportional to $g'$,
namely, $\delta H=K_1+K_2$.

The linear response of a Heisenberg operator $A(t)$, evaluated at $t=0$, to a perturbation in the Hamiltonian
of $\delta H$ is given by,
\begin{equation}
\delta A(0)=-i \int_{-\infty}^0 dt[A(0),\delta H(t)].
\end{equation}
Thus we can exhibit the term that is second order in $g'$ in the expression for the 
\underline {
operator} $dQ/dt$, as, 
\begin{equation}
dQ/dt=\int_{-\infty}^0 dt \Bigr [\Bigr (K_1(0)-K_2(0)\Bigr ),\Bigr (K_1(t)+K_2(t)\Bigr ) \Bigr ].
\label{linear}
\end{equation}
Following the steps of  Ref. \cite{bs}, appendix A, and noting that the terms involving two $K_1$'s or two $K_2$'s in (\ref{linear})
each give vanishing contributions, we use time translational invariance and the antisymmetry of the
commutator to obtain
\begin{equation}
w \equiv \langle dQ/dt \rangle=\int_{-\infty}^\infty dt \langle [K_1(t), K_2(0)] \rangle,
\label{wdef}
\end{equation}
where we have now used the brackets to indicate the expection in whatever ensemble we choose for the  $N,\Phi,L$
particles. We could also have obtained (\ref{wdef})
starting from the standard ``reduction formulae" of field theory.

Choosing an equilibrium ensemble for $N,\Phi,L$
we get the rate expression,

\begin{equation}
w=  Z^{-1} {\rm Tr}  \Bigr [e^{-\beta (H - \sum n_i \mu_i)} \int_{-\infty}^\infty dt~[K_1(t),K_2(0)] \Bigr ],
\label{formal}
\end{equation}
where Z is the partition function and where we  take the chemical potentials to satisfy the equilibrium condition $\mu_N=\mu_L+\mu_\Phi$. 
We specialize to the case in which there are no $\bar L$'s in the medium so that only the term with $K_1$ on the left in the commutator in (\ref{formal}) enters, and we set the expectation of $\bar c_i \bar c_i^\dagger$ equal to unity, giving,
\begin{equation}
w=g'^2\int dE_j \,f(E_j)\,r(E_j)\,,
\label{rate2}
\end{equation}
where
\begin{equation}
r(E)=Z^{-1} {\rm Tr}  \Bigr [ e^{-\beta( H -\sum n_i \mu_i)} \int_{-\infty}^\infty dt \,\,e^{i E t } [N(t)  \phi ^\dagger (t)]_j [N^\dagger (0) \phi (0)]_j \Bigr ].
\label{bs2}
\end{equation}

We note that the RHS of (\ref{bs2}) appears to depend on the index $j$. It is the independence of the $N$ and $\Phi$ energies on the respective particle momenta that makes the expression $j$ independent, after the trace has been taken. With our normalizations all
of the integrals over momenta in the perturbation expansion of (\ref{bs2}) will be unity, and we need concern ourselves with
the energy integrals only.
 
We now use the methods of thermal field theory to evaluate the average in (\ref{bs2}). In thermal field theory, however, a Heisenberg picture operator is defined by 
\begin{equation}
O(t)=\exp [i(H - \sum n_i \mu_i)t] \,\,O_{Sch}\,\,\exp [-i(H - \sum n_i \mu_i)t] ,
\label{heis}
\end{equation} 
and in (\ref{bs2}) the time dependence of  the operators $N(t)$ and $\phi^\dagger (t)$  is given by the Hamiltonian
alone. We can, however, use the time dependences of (\ref{heis}) in (\ref{bs2}) 
provided that $E$ in (\ref{bs2}) is evaluated in the end at $E=E_j-\mu_N+\mu_\phi=E_j-\mu_L$, rather than at $\omega_j$. We thus replace (\ref{rate2}) by

\begin{equation}
w=g'^2 \int dE_j\,  f(E_j)\,  r(E_j-\mu_L),
\label{finalrate}
\end{equation}
with the additional understanding that now the evaluation 
of $r(E)$ in (\ref{bs2}) proceeds according to the rules of equilibrium statistical mechanics.

As a point of terminology, we note that the operators that enter in the definition of the function $r(E)$, namely
$N \Phi^\dagger$ and $N^\dagger \Phi$, are the ``currents" that enter the RHS of the field equations for
the fields $L$ and $L^\dagger$, divided by $g$. Thus the thermal average of the unordered product, as we find it in (\ref{bs2}),
 is  directly related to the imaginary part of the self-energy function for the $L$ field. The construction
of thermal Greens functions proceeds in imaginary time, and thermal factors arise in the continuation back
to real energies. The result is

\begin{equation}
r(E)={2 \over 1+e^{\beta E}} {\rm Im}\Bigr [M (E+ i \eta)  \Bigr],
\label{rate3}
\end{equation}
where $M(E)$ is the imaginary time self-energy (or ``polarization part") for the $\bar L$, and $\eta$ is
infinitesimal. \footnote{The derivation of
(\ref{rate3}) for the case of fermionic currents is given in ref \cite{bs}, eq.s 4.3 to 4.8. It can also be found
in ref \cite{fw}, ch. 9, where it is, however, given explicitly for the Bose case, so that the appropriate change must be made
in the multiplicative thermal factor in order to obtain (\ref{rate3}). } The argument of $M_0$ reflects the continuation described above.
 Note that the imaginary part of this polarization part is not related to any real decay of the $L$, which
is stable; it arises from energy-conserving processes in which the 
$L$ meets a $\Phi$ from the medium, making whatever state, before coming back to the original configuration.

In the first part of the calculation of $M(E)$, given below,
$E$ is assigned a value that is equal to $\pi i m/\beta$ where $m$ is an odd positive or negative integer, according  to the
rule for the discrete energies of Fermions. At the end of this calculation we have a function of $E$ which despite having been defined on a discrete set of points is the unique extension to the whole $E$ plane that vanishes at infinite $E$. This function is to be continued to a nearly real energy, as indicated in (\ref{rate3}),
with the usual infinitesimal imaginary part in E. The quantities $\epsilon_\Phi=E_\Phi-\mu_\Phi$ and $\epsilon_N=E_N-\mu_N$
remain real throughout this calculation.
The lowest order single-loop graph for $L\rightarrow L$ has a ($N,\Phi$) loop with circulating energy, $\omega_n$ which we take to
be the energy of the $\Phi$ line. The inverse propagator for the $\Phi$ line is $\omega_n-\epsilon_\Phi$; then the energy associated with the
$N$ line is $E+\omega_n$ with inverse propagator $E+\omega_n-\epsilon_N$. The one-loop form for
$M$, which we denote by $M_0$ and which is of zero'th order in $g^2$, is given by the sum over Matsubara frequencies as,

\begin{equation}
M_0(E)=\beta^{-1} \sum_{n=-\infty}^{n=\infty}{1 \over (\omega_n - \epsilon_\phi )(E+ \omega_n- \epsilon _N)},
\label{me}
\end{equation}
where $\omega_n=2\pi i n /\beta$, in view of our choosing $\omega_n$ to be the energy of the intermediate
$\Phi$ (i.e. boson) line in the self-energy bubble.
This sum can be evaluated using, 
\begin{equation}
\beta^{-1} \sum_{n=-\infty}^{n=\infty}F(2\pi n i )=\int_{\rm C}{d \omega \over 2\pi i}{1 \over 2}
\cot \Bigr ({ \beta \omega \over 2})F(i \omega),
\label{whatever}
\end{equation}
which holds for a function $F(z)$ that has no singularities on the imaginary axis. 
The contour, C, runs from $-\infty$ to $\infty$ just below the real axis, then returns to $-\infty$ just
above the real axis. We get,

\begin{equation} 
M_0(E)=-\int_{\rm C}{d \omega \over 2\pi i}{1 \over 2}\cot \Bigr ({ \beta \omega \over 2})
\Bigr [ {1 \over (\omega -i \epsilon_\phi )(iE+ \omega-i \epsilon _N)}\Bigr ].
\label{me2}
\end{equation}
For evaluation on the lower part of the contour we use,
\begin{equation}
{1 \over 2 i} \cot \Bigr ({\beta \omega \over 2} \Bigr )={1  \over 2}+{1\over e^{i\beta \omega}-1}.
\label{iden1}
\end{equation}
For the upper part of the contour we use,
\begin{equation}
{1 \over 2 i} \cot \Bigr ({\beta \omega \over 2} \Bigr )=-{1 \over 2}-{1\over e^{-i\beta \omega}-1}.
\label{iden2}
\end{equation}

The contribution to $M_0 (E)$  from the $\pm 1/2$ terms in the identities (\ref{iden1}), (\ref{iden2}) can easily
be seen to vanish, since both poles of the multiplying factors of the integrand are in the upper-half plane.
The remaining parts, coming from the second terms on the RHS's of (\ref{iden1}) and (\ref{iden2}), are evaluated by noting that we can close the contours at infinity and calculate the residues at $\omega =i \epsilon_\phi $ and at $ \omega=-iE+i \epsilon _N$
(remembering that $E$ is imaginary whereas $\epsilon_N$ and $\epsilon_\phi$ are real and positive), giving,

\begin{equation}
M_0(E)={1 \over E-\epsilon_N+\epsilon_\phi }\Bigr [ {1\over e^{\beta \epsilon_\phi}-1}+{1\over e^{\beta \epsilon _N}+1}\Bigr ],
\label{m1}
\end{equation}
Note that the discrete value of the energy , $E=i\pi n/\beta$ with $n$ odd, entered here to create the Fermi distribution deriving
from the pole at $ \omega=-i E+i \epsilon _N$.
Next we use (\ref{m1}) in (\ref{rate3}), obtaining,
\begin{eqnarray}
r(E) =&2 \pi g'^2 \delta(E-\epsilon_N+\epsilon_\Phi) (1+e^{\beta \epsilon_N-\beta\epsilon_\phi})^{-1}
[ ( e^{\beta \epsilon_\phi}-1)^{-1}+(e^{\beta \epsilon _N}+1)^{-1}]
\nonumber\\
&= 2 \pi g'^2 \delta (E-\epsilon_N+\epsilon_\Phi) [ 1-e^{-\beta \epsilon_\Phi}]^{-1}[1+e^{\beta \epsilon _N}]^{-1},
\label{rzero}
\end{eqnarray}
leading finally to the rate, as given by (\ref{finalrate})
 \begin{equation}
w^{(0)}=2 \pi g'^2\int dE_j \, f(E_j) \delta(E_j+E_\Phi-E_N)  [ 1-e^{-\beta \epsilon_\phi}]^{-1}[1+e^{\beta \epsilon _N}]^{-1}.
\label{gr}
\end{equation}
Eq. (\ref{gr}) is just the golden rule (\ref{w0})

for the lowest order $N \rightarrow \bar L +\Phi$ process with the appropriate thermal factors added for the $N$ and the $\Phi$.

While it may seem to have been a pointless task to derive the trivial lowest order
rate in the above complicated way, the reward is that the corrections to order $g^2$ are now relatively easy to define and to calculate.
We need only to put in the correction to the N inverse propagator,
replacing
 $E+\omega_n -\epsilon_N$ in  (\ref{me}) by $E+\omega_n -\epsilon_N-\Pi(E+\omega_n)$, where,
\begin{equation}
 \Pi(x)=g^2 \int dE_j  f(E_j) \Bigr [\sum_{n'=-\infty}^\infty{1 \over \omega_{n'}-\epsilon_\phi}\,\,{1 \over x -\omega_{n'} -\epsilon_j}
\Bigr],
\end{equation}
and where $\epsilon_j=E_j-\mu_L$. The function $\Pi$ is the self-energy bubble in which the N goes virtually into a $\Phi$ and an $L$.

There is a completely analogous correction to the $\Phi$ inverse propagator, which we will not write down. If we made the appropriate replacement of $\omega_n-\epsilon_\Phi$ in (\ref{me}) and expanded, we would generate the 
scattering process B, i.e. $N+L \rightarrow \Phi +\bar L+L \rightarrow N+ \bar L$, \underline{plus} the appropriate
singularity-regulating corrections to the basic decay $N\rightarrow \bar L + \Phi$. The results of the calculations are completely
parallel to those we will discuss that relate to process A. For simplicity only, we stick to the range of parameters
where the B channel is disfavored.

In the language being used in the literature about this subject area, the relevant graphs
have already been ``resummed" when $\Pi$ is included in the inverse propagator. Indeed, the papers 
to which we compare our results keep only ${\rm Im}[\Pi(x)]$, where $x=\epsilon_N$ and the appropriate continuation
has been made before taking the imaginary part.
However, as we see below, no ``resumming'' of any kind is needed in order
to get the complete and finite answer to order $g^2$. Thus we simply use only the order $g^2$
term in the expansion of  the propagator, i.e. we make the replacement in (\ref{me}),

\begin{equation}
(E+\omega -\epsilon_N)^{-1} \rightarrow (E+\omega -\epsilon_N)^{-1}+ (E+\omega -\epsilon_N)^{-2} \Pi(E+\omega ).
\label{trick}
\end{equation}
The last term supplies the order $g^2$ correction to the rate, given simply by replacing $M_0$ in (\ref{rate3}) by
\begin{eqnarray}
M_1(E)=g^2\beta^{-1} \sum_{n=-\infty}^{n=\infty}\Bigr[{1 \over \omega_n - \epsilon_\Phi }{1 \over (E+ \omega_n- \epsilon _N)^2}
\nonumber\\
\times \int dE_j \,f(E_j) \,
\sum_{n'=-\infty}^{n'=\infty}{1 \over \omega_{n'}-\epsilon_\Phi }\,\,{1 \over E+\omega_n -\omega_{n'} -E_j}\Bigr],
\label{m11}
\end{eqnarray}
and then substituting in (\ref{finalrate}).

Actually, the structure of the expression (\ref{m11}) probably is sufficient by itself to show that the answer is
defined, and finite, although we will show below the explicit form that results from doing the 
$\omega_n$, $\omega_{n'}$ sums, again using the identities (\ref{iden1}) and (\ref{iden2}), followed be the continuation
to (almost) real values of the energy, followed by taking the imaginary part. Indeed, the problem
with the perturbation theoretic S-matrix approach that led to the perceived, and spurious, need for the ``resummation", is precisely
the double pole in the integrand, $(E+ \omega_n- \epsilon _N)^{-2}$. In the S matrix approach 
(in the superficial treatments that have been followed) a closely related factor appears under an integral in
a generic form $(E+ \omega- E _N+i\eta)^{-1}(E+ \omega- E _N-i\eta)^{-1}$, where $\eta$ is infitesimal,
the integration contour being squeezed between the singularities on both sides. 

In (\ref{m11}), no such
problem arises in the evaluation. One way of seeing this, and the best first step before doing the 
frequency sums, is to write $(E+ \omega_n- \epsilon _N)^{-2}=(\partial /\partial \epsilon_N)(E+ \omega_n- \epsilon _N)^{-1}$.
\footnote{This trivial trick is unavailable in the standard S matrix approach if we have resolved into exclusive final states
since in that case the denominator function is not a square, but rather an absolute square.} Then the sums are performed using (\ref{whatever}), followed by the continuation, taking the imaginary part, and finally taking the 
$\epsilon_N$ (or $E_n$) derivative. Doing the frequency sums, first for the inner sum in (\ref{m11}), and then for
the outer sum, is a straightforward mechanical procedure, and we omit the details. The answer, after we substitute
in (\ref{finalrate}) is remarkably simple,
\begin{eqnarray}
w^{(1)} =2 \pi g'^2 g^2 {1 \over 1-e^{-\beta (E_\phi -\mu_ \phi)}}\,\,{1 \over e^{\beta (E_\phi- \mu_\phi)}-1}
\int dE_j \,  {f(E_j) \over 1+e^{\beta (E_j-\mu_L)}}
\nonumber\\
\times {\partial \over \partial E_N}\Bigr [ [{ 1\over E_j+E_\phi-E_N}][f(E_j)-f(E_N-E_\phi)\,{1+e^{\beta(E_j+E_\phi-\mu_N)} \over 1+e^{\beta (E_N-\mu_N)}}] \Bigr ].
\label{preanswer}
\end{eqnarray}
This is the complete second order correction to the rate for $\bar L$ appearance in the model
described by the original Hamiltonian. We note that the term with $f(E_j)$ in the last factor is exactly
the (divergent) rate calculated from the $L+\Phi \rightarrow L + \Phi$ Feynman graph, without ``resumming".

Before interpreting the other term, which removes the divergence, we take the limit of Boltzmann statistics (all $\mu$'s large and negative) 
\begin{eqnarray}
&w^{(1)}_B=g^2 g'^2 e^{\beta (\mu_\phi+\mu_L-E_\Phi)}
\int dE_j \,  f(E_j) e^{-\beta E_j}
\nonumber\\
\times &{\partial \over \partial E_N}\Bigr [ [{ 1\over E_j +E_\phi-E_N}][f(E_j)-f(E_N-E_\phi)e^{\beta(E_j+E_\phi-E_N)}] \Bigr ].
\label{answer}
\end{eqnarray}
The integrands (after taking the continuum limit) in both (\ref{answer}) and in (\ref{preanswer}) are perfectly
finite; they do not require regulation, or even principal value prescriptions. Moreover, there is no assurance that $w^{(1)}$ is positive; the counter term with $f(E_N-E_\Phi)$ can easily
outweigh the first term in the last factor in (\ref{answer}), and does so for some simple choices of $f(E)$.

\section{Alternative approach} 
There are two other ways of obtaining the same results. 
In a careful real-time treatment as well, the inclusive rate
$w$ can be expressed as the imaginary part of a two-loop amplitude, where the propagators are standard real
time finite temperature propagators. In a correct formulation (with $i\eta $'s defining the integrals) all of the 
singularities are on the same side of the integration contours, so that this amplitude is non-singular.  The real time 
approach may be less authoritative even if is a quicker
way to get to the result. See ref. \cite{kapusta} for a demonstration of its delicacy in some circumstances
similar to the ones of this paper. A real time approach based on Keldysh ($2\times 2$ matrix) Greens functions
should be problem-free however.\footnote{In principle, the Keldysh Greens function approach
of ref.\cite{bf} should produce our results, both in the toy
model and in the genuine leptogenesis models. The particular
contribution that we have discussed is contained in the 
two loop graph of fig. 12c of this paper, a self energy
part for the $l$ fields. However in the sample evaluation given in ref.\cite{bf}
of the imaginary part of this graph, the internal $N$ propagator 
is taken as an energy-independent (and point) interaction, 
so that none of the problems that we have addressed are present. 
If instead one uses the propagator that creates the singularities
we have discussed in this paper, there will appear to be 
squares of delta functions in individual terms, which are characteristic
of real time approaches. But in the matrix (Keldysh) approach
of ref.\cite{bf} these terms should cancel, once again, so that
there is no need for ``resumming" for purposes of regulation.}

The third  approach goes back to the vacuum S matrix calculation based on the Feynman rules but
treats in addition the medium-dependent wave-function renormalization (MDWFR) that is required
in such an approach \cite{mdwfr}. 
We begin from the lowest order result for the Boltzmann statistics case, but modified as follows,
\begin{equation}
w^{(0)}_B=2 \pi g'^2 e^{\beta (\mu_N-E_N-\Delta)} f(E_N-E_\Phi+\Delta)Z_2,
\end{equation} 
which is just the Boltzmann limit of (\ref{w0}) but with an energy shift $\Delta$ for the $N$
particle and a factor $Z_2$, which is a redefinition of the coupling constant. Now we calculate the
lowest order difference between this modified rate and the earlier lowest order result,

\begin{eqnarray}
\delta w ^{(0)}_B= g'^2 e^{\beta (\mu_N-E_N)}
 \Bigr [ \Delta   \Bigr(  -\beta +{\partial \over \partial E_N} \Bigr ) f(E_N-E_\Phi )  
+(Z_2-1) f(E_N- E_\Phi ) \Bigr ].
\label{modified}
\end{eqnarray} 

The $N$ propagator, with the single bubble self energy part is,
\begin{equation}
S(E)=\Bigr [E-E_N+g^2\int dE_j \, {f(E_j)\over E_j+E_\Phi-E} \Bigr ]^{-1}.
\end{equation} 
The identification of the energy shift $\Delta$ and $Z-1$ is accomplished by calculating the
position of the pole of the propagator and the associated residue, both to lowest order \footnote{These MDWFR
parameters are not even dependent on the medium, in a sense, in the present calculation. Note, however, that they
would have been, if we had used the statistical factors in (\ref{preanswer}) instead of taking the Boltzmann limit.},
obtaining,
\begin{eqnarray}
\Delta=-g^2 \int dE_j {f(E_j) \over E_j+E_\Phi-E_N},
\label{delta}
\end{eqnarray}
and
\begin{eqnarray}
Z_2-1=[ 1+g^2 \int dE_j {f(E_j)\over  (E_j+E_\Phi-E_N)^2}]^{-1}-1\approx - g^2 \sum_k {f(E_j)\over  (E_j+E_\Phi-E_N)^2}.
\label{renorm}
\end{eqnarray}
Now we see that we can replace our imaginary time calculation of the last section by adding the ``Feynman graph"
result, the first term on the RHS of (\ref{answer}), to $\delta w^{(0)}$ from (\ref{modified}), using (\ref{renorm}) and (\ref{delta}), and defining 
the result as $ w_{mdwfr}^{(1)}$. The difference between this result and (\ref{answer}) is given by, 
\begin{eqnarray}
w^{(1)}_B-w_{mdwfr}^{(1)}=g^2 g'^2 e^{\beta  \mu_N}
\int dE_j  f(E_j) \Bigr \{{\partial \over \partial E_N}\Bigr [ { f(E_N-E_\phi)e^{-\beta E_N)}\over E_j+E_\Phi -E_N}  \Bigr ]
\nonumber\\
-{e^{-\beta E_N}[-\beta +{\partial \over \partial E_N} f(E_N-E_\Phi)]\over E_j +E_\Phi -E_N}-{e^{- \beta E_N}f(E_N-E_\Phi)\over (E_j +E_\Phi-E_N)^2}\Bigr \}=0.
\end{eqnarray}
 Thus we can get the identical result as (\ref{answer}), with a lot less work in the case of the present model, using
 the methods of this section.
There are some reasons to favor the imaginary time approach of the previous section :
a.) Rather than depending on the cancellation of terms that are only defined formally, in a sense,
the answer is never separated into individually divergent terms.
b.) In the model at hand, it is the $N$ propagator that serves to define the renormalization constants $\Delta$ and
$Z_2$. Since $N$ is an unstable particle, these definitions could be questioned.
c.) In the more difficult relativistic problem studied in \cite{bs} it was found that the terms $\pm 1/2$ in (\ref{iden1}) 
and (\ref{iden2}) played the special role of picking out the ``vacuum" part of the expression, a part that
was independent of the temperature and the density of the medium, but that carries ultraviolet divergences
that are to be addressed in the usual fashion.

\section{Discussion}

We explain further why the toy model poses the right question: 

a. \underline {Momenta and relativity}. We not only left out
relativity, but we discarded the kinetic energies of two of our four kinds of particles. 
In a limit in which the $N$ and $\Phi$ are very massive, this is not such a bad mutilation, since
the kinetic energy \underline{change} for the heavy particle in any of our reactions would be small
in any case. What if all the particles had finite mass and were relativistic? In the first place we would
have a bigger manifold of non-trivial integration for the two loop processes. In (\ref{m11}) we had a sum
over momenta for the $L$ line only; in the realistic case there will be integrals over two three-dimensional
loop momenta, and the constants $E_N,E_\Phi$ would be replaced by functions of the momenta. Moreover,
there would be many further dependences of the summand on the loop energies, re-expressed as
Matsubara frequencies, coming from the relativistic propagators. However, as regards to the 
double-pole singularity in energy which enters, when the innocent factors multiplying this pole are evaluated
at the pole position, the resulting structure will recapitulate \underline{exactly} the structure
that we dealt with above. Thus we are assured that there is no ``resummation" required to obtain
a finite answer, again of order $g'^2 g^2$, and therefore probably no real point in doing the 
detailed calculation.

b. \underline{The initial state.} The actual models being used in leptogenesis are of sufficient internal complexity 
to rather frustrate the non-specialist (such as the present author) in confronting some of the details
in this regard. Thus we avoid any comments on the larger issues and stick with our model with
four species. But to begin with our first arbitrary assumption, that the number of $\bar L$'s in the medium
was zero, this was just for convenience, in order to have half as many terms to deal with. When, in the derivation
of (\ref{rate2}) we set $\bar c_i \bar c_i^\dagger=1$, and $\bar c_i \bar c_i^\dagger=0$ we could have set

\begin{eqnarray}
\langle \bar c_i \bar c_i^\dagger\rangle =1-n_F(E_i-\mu_{\bar L}),
\nonumber\\
\langle \bar c_i ^\dagger \bar c_i\rangle =n_F(E_i-\mu_{\bar L}),
\end{eqnarray}
and calculated separately the terms for the creation and destruction of $\bar L$'s, and, if we liked, the back reaction
on $\mu_{\bar L}$. Taking the field $\bar L$ to be the true antiparticle
manifestation of the field $L$ and adding an equilibration process,
(self-conjugate boson) $ \leftrightarrow L +\bar L$, we could also watch the system move towards its
true equilibrium with equal numbers of $L$ and $\bar L$ particles and $\mu_L=\bar \mu_L=0$.
But our point was rather to show how we can calculate the separate rates for the terms that
increase and decrease the lepton number, and to show that these terms are individually free
of the ``resonance" pathology. Could one contrive an initial state in which the pathology was real?
Perhaps one could, although the answer is not obvious. But, to put questions like this into context,
we might reflect on the role of statistical mechanics in early universe problems. To the extent
that we have equilibrium, everything is well defined and it is easy to retrieve the results.
We can also address situations that are out of equilibrium
by a little, or which are in thermal but not chemical equilibrium. The tools that we have described
are sufficient for these cases. To go farther in taking systems out of equilibrium is tricky. We must hope that physics does not force us to follow the complete history of each degree of freedom.

We note that the authors of Ref. \cite{giudice} agree with the qualitative conclusion
(no resonant enhancement) that we express above. However, they achieve this result by
subtracting twice as much from the ``resonant" term than is implied by the replacement (\ref{replacement2}).
Besides being arbitrary (in our opinion), this procedure leads to completely incorrect
residual terms of order $g^2 g'^2$ as one can see by comparison with our results. If one wants to follow
the graph-based (exclusive) approach, there appears to be \underline{no alternative} to
correctly treating the wave function renormalization processes as in the last section. 

To calculate the analogue
of our $g^2 g'^2$ terms, but in a full relativistic model, one could follow the steps more or
less outlined above, first reducing the question to that of the equilibrium average of a product
of currents, then using the standard methods of statistical mechanics. We would argue, however, that
one never needs these terms. If $g$ is small, then they don't matter compared to the straight $g'^2$
term. If $g$ is large, perturbation theory is inapplicable, and the answer is anybody's guess.
Since we are not working in a situation in which coupling constants are exactly known, and
since we do not seek precision results, these higher order effects must be irrelevant.

We have followed a very computational path to arrive at our conclusions. There surely are
considerations that would obviate the need for ``toy-models" to demonstrate the existence
of the perturbation expansion of inclusive rates that derive from formal expressions like that of Eq. (\ref{bs2}),
even when the expansions for exclusive rates do not exist. We reemphasize that that is exactly what 
happened in the toy calculation; if we do exclusive calculations with graphs, the corrections 
to the $N \rightarrow \Phi  +\bar L $ rate and the lowest order $\Phi+L \rightarrow  \Phi+\bar L$,
each separately contain terms that are infinite in perturbation theory (without the $i \gamma$ device),
but which cancel in the inclusive rate. A graphical calculus of the inclusive rate begins with
a current at one end and another current at the other end, with instructions to take the imaginary
part to get the rate. When graph has two or more loops \footnote{Note that in our terminology, the basic $N$ decay
is ``one-loop" and the scattering process is ``two-loop"}, then there are ``cutting rules" which
in principle give the inclusive rate, but where the individual cuts give the contributions of
various processes. These rules, as well as the general field-theory background for our calculations are given in Ref. \cite{weldon}. What we have shown is that it is advisable not to use these ``cutting rules", but
instead to do some integrals and combine some terms before taking the imaginary part. There is probably an provable general result that the inclusive rate is free
of the pathologies addressed in this paper. In an analogous case of radiative processes in
zero-temperature field theory, this was proved long ago \cite{kinoshita}-\cite{LN}.
Thus, for example, bremsstrahlung divergences in exclusive amplitudes do not affect one's ability to compare laboratory cross-sections to completely perturbative predictions. 

The reader may note that in the present model the problems arise because of an unstable field,
$N$. What if we change the masses so that the decay is forbidden? Then, of course, to the
order that we have considered, there are no issues to address. If one were rash enough to look at perturbative corrections
to these rates, however, involving more particles in the medium as well as corrections that
would be called ``radiative", all of the same issues will surface. Following the S matrix approach
(adding up graphs and weighting with thermal factors) in this case might well give finite correction terms.
But they will be the wrong terms, unless we do (finite) medium-dependent wave-function renormalization
correctly. The issues raised in this paper are really not those of treatment
of unstable particles, but rather those of how one does field theory in a medium.

To sum up with respect to the leptogenesis models: in our opinion there will not be a resonant enhancement
of an inclusive rate for change in lepton number. If our toy model is deemed inadequate for making
this point, then the author making this claim should at least find an alternative to the incorrect
procedures of Eqs. (\ref{replacement1}) and (\ref{replacement2}); if these procedures are conceptually
wrong in a simple case, then they are wrong in a more complicated case. \footnote{We should emphasize
that our remarks have no bearing (that we know of) on the treatement of models with
a number of Fermion species, closely spaced in mass, for example in Ref. \cite{akmedov}.}

We comment briefly on another problem in which exactly the same issues have surfaced, and
in which ``resumming", in the sense used above, also gives correction
terms that are both of the wrong order in coupling parameters and of the wrong sign. The
physical situation is the scattering of a neutrino from a nucleon in the supernova environment.
There is a ``zero'th order" quasielastic rate which is just the (neutral current) vacuum rate, 
given by cross-section times density. But if one looks at a Feynman graph for the correction
terms in which the nucleon that interacts with the neutrino is scattered by other nucleons,
the (squared) first order correction to the amplitude has a singularity that is non-integrable, 
this coming from the kinematical region in which no energy is exchanged. The authors 
of Ref.s \cite{r1}-\cite{r4} have addressed this problem by
putting a finite imaginary part in a denominator, and relating this imaginary part to a (strong)
nucleon-nucleon scattering rate in the medium. 

But in a consistent perturbation development the
singularity in question is cancelled  by the interference between the second order amplitude,
dead forward and elastic for the spectator particle, and the zeroth order result, disconnected
from the spectator \cite{rfs}. The interpretation of this last term is exactly that of medium-dependent
wave function renormalization. 
We emphasize that while doing calculations this way,
beginning with Feynman graphs, can get to the right answer if done carefully enough and in
a framework that recognizes wave-function renormalization, the {\it ab initio} methods
of the present paper are a more convincing way of getting the same results. Interestingly,
although the conventional name given to the source of the difficulties in the neutrino problem
(a ``colinear divergence") is an entirely different name to that given to the source in the leptogenesis problem
(``unstable field"), the resolution in the two cases is almost identical.

This neutrino
problem can also serve to crystallize one's understanding of why the effect of adding an interaction that
creates a new channel for producing something often gives a reduction in the rate of this production. In every such problem there is a sum rule that
says that the integrals of momentum-space matrix elements of some weak currents, weighted by
kinematical factors, are independent of the other interactions in the medium. When we turn
on, say, the nucleon-nucleon interaction in the above example, the ``strength" as defined
by the sum rule, is rearranged. The strength often moves toward domains that are out of the kinematical
domain allowed for the process that we are calculating. Thus we are not surprised that even
when an inelastic channel is opened for neutrino scattering, by including $N$-$N$ interactions,
 more than enough strength is taken away
from the zero'th order (quasielastic) channel to reduce the net neutrino scattering rate. In the problem discussed
in the present paper the sign of the first order correction depends on the function $f(E_k)$. In the actual leptogenesis
calculation, done correctly, it could well be that the first order changes would be a reduction of the reaction rate, not
the increase that one finds in the literature.

We cannot refrain from commenting on an irony of history. In the late 1960's in the particle physics world it was politically
incorrect to express the belief that there is an underlying Hamiltonian. So people asked, ``if you don't have a
Hamiltonian, how do you find the partition function, Tr$[\exp(-\beta H)]$?". The classic paper 
by Dashen, Ma and Bernstein \cite{dashen} answered exactly this question. This paper showed very
generally how one can find the partition function from S matrix elements, for the case of particles interacting via two-body
interactions. Of course, our questions go a little farther than that of finding a partition
function, and the interaction in our problem is more complex as well, since we have particle production and annihilation. If we ask, ``how do we answer our questions using only S matrix elements?" (which is surely
possible), the answer must be ``very carefully". The irony is that in the present era we work from a Hamiltonian, 
and that it is intrinsically clearer just to calculate the quantities of interest from this Hamiltonian
directly, rather than going through the intermediary of an S matrix. 

\section{Acknowledgements}
I am indebted to Yvonne Wong for bringing this literature to my attention during
the Spring 2003 workshop on neutrino physics at the Kavli Institute for Theoretical
Physics, and to Lowell Brown for indoctrination in the techniques used in this paper.

\end{document}